
%

\overfullrule=0pt
\magnification=\magstep1
\newdimen\theight
\def \Column{%
             \vadjust{\setbox0=\hbox{\sevenrm\quad\quad tcol}%
             \theight=\ht0
             \advance\theight by \dp0    \advance\theight by \lineskip
             \kern -\theight \vbox to \theight{\rightline{\rlap{\box0}}%
             \vss}%
             }}%
\catcode`\@=11
\def\qed{\ifhmode\unskip\nobreak\fi\ifmmode\ifinner\else\hskip5\p@\fi\fi
 \hbox{\hskip5\p@\vrule width4\p@ height6\p@ depth1.5\p@\hskip\p@}}
\catcode`@=12 


\newcount\notenumber
\def\clearnotenumber{\notenumber=0}
\def\note{\global\advance\notenumber by 1
 \footnote{$^{\the\notenumber}$}}
\clearnotenumber
\def\mapright#1{\smash{\mathop{\longrightarrow}\limits^{#1}}}

\def\mapdown#1{\Big\downarrow\rlap{$\vcenter{\hbox{$\scriptstyle#1$}}$}}

\medskip
\rightline{hep-th/9303111}
\bigskip
\centerline{{\bf Dilogarithm Identities in Conformal Field Theory and
Group Homology}$^{*)}$}
\bigskip

\centerline{Johan L. Dupont}
\centerline{Matematisk Institut}
\centerline{Aarhus Universitet}
\centerline{Ny Munkgade, DK-8000 $C$}
\centerline{Aarhus, Denmark}
\bigskip
\centerline{and}
\bigskip
\centerline{Chih-Han Sah}
\centerline{Department of Mathematics}
\centerline{SUNY at Stony Brook}
\centerline{New York 11794-3651, USA}
\vskip24pt plus8pt minus8pt
{\bf Abstract.}  Recently, Rogers' dilogarithm identities have
attracted much attention in the setting of conformal field theory as well as
lattice model calculations.  One of the connecting threads is an identity of
Richmond-Szekeres that appeared in the computation of central charges in
conformal field theory.  We show that the Richmond-Szekeres identity and its
extension by Kirillov-Reshetikhin can be interpreted as a lift of a generator
of the third integral homology of a finite cyclic subgroup sitting inside the
projective special linear group of all $2 \times  2$ real matrices viewed as a
{\it discrete} group.  This connection allows us to clarify a few of the
assertions and conjectures stated in the work of Nahm-Recknagel-Terhoven
concerning the role of algebraic $K$-theory and Thurston's program on
hyperbolic 3-manifolds.  Specifically, it is not related to hyperbolic
3-manifolds as suggested but is more appropriately related to the group
manifold of the universal covering group of the projective special linear group
of all $2 \times  2$ real matrices viewed as a topological group.  This also
resolves the weaker version of the conjecture as formulated by Kirillov.
We end with the summary of a number of open conjectures on the mathematical
side.
\vskip60pt plus20pt minus20pt
\noindent
$\_\_\_\_\_\_\_\_\_\_\_\_\_$

\noindent
$^{*)}$This work was partially supported by grants from Statens
Naturvidenskabelige Forskningsraad, and the Paul and
Gabriella Rosenbaum  Foundation.

\vfill\eject

{\bf {\S}0.  Introduction.}
\medskip
Very recently, much has been written about the Rogers' dilogarithm
identities and its role in conformal field theory, see [BR], [KKMM], [FS],
[K], [KR], [KP], [KN], [KNS], [NRT].  For an excellent general survey for
mathematicians concerning hypergeometric functions, algebraic $K$-theory,
algebraic geometry and conformal field theory, see [V] and its extensive
section of references.  For a recent review from the physics side, see
[DKKMM].  In the present work, we limit our attention to the special case
of dilogarithm identities.  In spirit, it fits into the program surveyed
by Varchenko [V].  Some, though not all, of the relevant calculations
have been carried out on both sides of the fence.  Conjectures abound
even in this case.  Most of our task consists of pulling together items
that are scattered in the literature invarious forms.  The new ingredient
is to give a direct interpretation in terms
of group homology to account for the Richmond-Szekeres identity, see [RS],
and its extension by Kirillov-Reshetikhin, see [KR, II, (2.33) and
Appendix 2].  What we show is that the basic identities are the ones found by
Rogers in [R].  Rogers' dilogarithm function then leads to a real valued
cohomology class defined on the third integral homology of the universal
covering group of $PSL(2,R)$, viewed as a {\it discrete} group.  The
Richmond-Szekeres identities, see [RS], and the Kirillov-Reshetikhin
identities, see [KR II, (2.33) and Appendix 2], are the results of
restricting the evaluation of this cohomology class (the real part of the
second Cheeger-Chern-Simons class) to the inverse image of a suitable homology
class that covered a generator of suitable finite cyclic subgroup.  This
will then provide partial clarifications of some of the assertions and
conjectures made by Nahm-Recknagel-Terhoven [NRT] related to algebraic
$K$-theory [Bl] and Thurston's program on hyperbolic 3-manifolds [Th2].
Specifically, we show that it is more appropriately related to the group
manifold underlying the universal covering group of $PSL(2,R).$
\medskip
{\bf {\S}1.  Rogers' Dilogarithm.}
\medskip
Rogers' dilogarithm (also called Rogers' $L$-function) was defined in $[R]:$

$$\eqalign {L(x) &= -{1 \over 2}\{\int ^x_0 {\log \ x \over 1-x} dx + \int ^x_0
{\log (1-x) \over x} dx\}. \cr
&= \sum _{n > 0}{x^n \over n^2} + {(\log \ x)\cdot (\log (1-x)) \over 2}\hbox{,
} 0 < x < 1. \cr
L(x) &\hbox{is real analytic, strictly increasing
and }\ \lim _{x\rightarrow 1}L(x) = \pi ^2/6. \cr } \leqno(1.1) $$
\medskip
\noindent
Rogers showed that $L$ satisfied the following two basic identities:

$$L(x) + L(1-x) = \pi ^2/6\hbox{, } 0 < x < 1. \leqno(1.2) $$

$$L(x) + L(y) = L(xy) + L({x-xy\over 1-xy}) + L({y-xy\over 1-xy})\hbox{, } 0 <
x\hbox{, } y < 1. \leqno(1.3) $$
\medskip
\noindent
If we use (1.2), take $s_1 = (1-x)/(1-xy)$ and $s_2 = y(1-x)/(1-xy)$ so that $y
= s_2/s_1$ and $x = (1-s_1)/(1-s_2)$ with $0 < s_2 < s_1 < 1$, then (1.3) is
seen to be equivalent to:
$$
L(s_1) - L(s_2) + L({s_2\over s_1}) - L({1-s^{-1}_1\over 1-s^{-1}_2}) +
L({1-s_1\over 1-s_2}) = {\pi ^2\over 6}, 0 < s_2 < s_1 < 1. \leqno(1.4)
$$
\noindent
If we set $r_i = s^{-1}_i$, then (1.4) can be rewritten in the form:
$$
L(r_1) - L(r_2) + L({r_2\over r_1}) - L({r_2-1\over r_1-1}) +
L({1-r^{-1}_2\over 1-r^{-1}_1}) = {\pi ^2\over 6}, 1 < r_1 < r_2. \leqno(1.5)
$$
\noindent
Motivated by [DS1], Rogers' dilogarithm was shifted in [PS] to:
$$
L^{PS}(x) = L(x) - {\pi ^2 \over 6} = -L(1-x)\hbox{, } 0 < x < 1. \leqno(1.6)
$$
\noindent
If we replace $L$ by $L^{PS}$ throughout, then (1.2) and (1.4) become:
$$L^{PS}(x) + L^{PS}(1-x) = - {pi ^2 \over 6}, \leqno(1.7) $$
$$L^{PS}(x) - L^{PS}(y) + L^{PS}({y\over x}) - L^{PS}({1-x^{-1}\over
1-y^{-1}}) + L^{PS}({1-x\over 1-y}) = 0, 0 < y < x < 1. \leqno(1.8) $$
\noindent
A huge number of identities have been found in connection with Rogers'
dilogarithm.  The situation is somewhat similar, and is often, related to
trigonometry, where the basic identities are the two additional formulae for
the sine and cosine function, which are just the coordinate description of the
group law for SO(2) or $U(1)$.  This analogy can be made more precise.  Namely,
$U(1)$, more appropriately, $GL(1,C) \cong  C^\times $ is just the first
Cheeger-Chern-Simons characteristic class in disguise.  This is wellknown and
tends to be overlooked.
\medskip
Richmond-Szekeres [RS] obtained the following identity (in a slightly different
form) from evaluating the coefficients of certain Rogers-Ramanujan partition
identities as generalized by Andrews-Gordon:
$$\sum _{1\leq i\leq r}L(d_i) = {\pi ^2 \over 6} \cdot {2r \over 2r+3},
d_j = {\rm sin^2 \theta \over \rm sin^2 (j+1)\theta}, \theta  = {\pi \over
2r+3}.  \leqno(1.9) $$
\noindent
This has been extended by Kirillov-Reshetikhin [KR] to:
$$\sum _{1\leq j\leq n-2}L(d_j) = {\pi ^2 \over 6} \cdot {3(n-2) \over n}
\hbox{, } d_j = {{\rm sin^2} \theta \over {\rm sin^2} (j+1)\theta} \hbox{, }
\theta  = {\pi \over n}. \leqno(1.10) $$
\noindent
Apparently, identity (1.9) arose in the study of low-temperature asymptotics of
entropy in the RSOS-models, see [ABF], [BR], and [KP] while (1.10) arose in the
calculation of magnetic susceptibility in the XXZ model at small magnetic
field, see [KR].  They are connected to conformal theory in terms of the
identification of the right hand sides as the effective central charges of the
non-unitary Virasoro minimal model and with the level $\ell \ A_1^{(1)} \ WZW$
model respectively, see [BPZ], [Z2], [K], [KN], [KNS], [DKKMM], [KKMM]
[Te],$\cdot \cdot \cdot $.  Our goal is to show that these identities can be
understood in terms of the evaluation of a Cheeger-Chern-Simons characteristic
class on a generator of the third integral homology of a finite cyclic group of
order $2r + 3$ and $n$ respectively.
\bigskip
{\bf {\S}2.  Geometry and algebra of volume calculations.}
\medskip
In any sort of volume computation, the volume is additive with respect to
division of the domain into a finite number of admissible pieces.  Depending on
the coordinates used to describe the domain the volume function must then
satisfy some sort of ``functional equation''.  This is the geometric content
behind the Rogers' dilogarithm identity.  The geometric aspect was
described in [D1] while some of the relevant algebraic manipulations were
carried out in [PS] (up to some sign factors that only became important in
[D1]).  To get a precise description, it is necessary to examine [DS1], [DPS]
and [Sa3].  These used algebraic $K$-theory.  We review the
ideas and results but omit the technical details.
\medskip
To begin the review, we recall the definition of some commutative groups
(called the ``scissors congruence groups'', cf. [DPS].  Let $F$ denote a
division ring (we are only interested in three classical cases:  $R = real$
number, $C = complex$ numbers, $H = quaternions.)$.  The abelian group $P_F$ is
generated by symbols:  $[x]$, $x$ in $F$, $x \neq  0, 1$ and satisfies the
following identity for $x \neq  y:$

$$[xyx^{-1}] = [y]\hbox{, (this is automatic for fields)} \leqno(2.1a)$$

$$[x] - [y ] + [x^{-1}y] - [(x-1)^{-1}(y-1)] + [(x^{-1}-1)^{-1}(y^{-1}-1)] = 0
\leqno(2.1b) $$
\medskip
\noindent
This group was studied in [DS1] for the case of $F = C$.  It is closely
related to, but not identical to, the Bloch group that was studied in [Bl].
A second abelian group $P(F)$ is defined by using generating symbols $[[x]]$,
$x$ in $F - \{0,1\}$, with defining relations:
$$\hbox{same as (2.1) with $[[z]]$ in place of $[z]$.} \leqno(2.2) $$

$$[[x]] + [[x^{-1}]] = 0. \leqno(2.3) $$

$$[[x]] + [[1-x]] = cons(F) \hbox{ (a constant depending on }\ F). \leqno(2.4)
$$
\noindent
The following result can be found in [DPS]:
$$0 \rightarrow  F^\times /(F^\times )^2 \rightarrow  P_F
\rightarrow  P(F) \rightarrow  0 \hbox{ is exact for } F = R \hbox {, } C \hbox
{, } H. \leqno(2.5) $$
\noindent
The first map in (2.5) is defined by sending $x$ in $F - \{0,1\}$ to $[x] +
[x^{-1}]$.  The second map then sends $[x]$ to $[[x]]$.  In particular, when $F
= C$, we may set $[x] = 0$ for $x = \infty $, 0, 1 and remove the restriction
$x \neq  y$ in (2.1) by adopting the convention:  meaningless symbols are taken
to be zero, see [DS1].  For the division ring $H$, we observe that every
element of $H$ is conjugate to an element of $C$, thus $P(H)$ is a quotient of
$P(C).$
\medskip
The geometric content of (2.1b) is best seen by thinking in terms of a
Euclidean picture.  Suppose we have 5 points in Euclidean 3-space so that
$p_1$, $p_2$, $p_3$ form a horizontal triangle while $p_0$, $p_4$ are
respectively above and below the triangle.  The convex closure is divided by
the triangle into two tetrahedra and also divided into three tetrahedra by the
line joining $p_0$ and $p_4$, see (Fig.1).

\bigskip
\centerline{(Fig. 1)}
\bigskip
\noindent
Thus, if any function of a tetrahedron is additive with respect to finite
decompositions, it would follow from (Fig. 1) that there should be a 5 term
identity to be satisfied by such a function.
\medskip
We examine the special case of $F = C$.  Here $P_C = P(C)$ is known to be a
$Q$-vector space of continuum dimension, see [DS1].  It is best to consider
the (-1)-eigenspace $P(C)^-$ of $P(C)$ under the action of complex
conjugation. It is classically known that the projective line $P^1(C)$ can be
viewed as the boundary of the hyperbolic 3-space.  An ordered set of 4 n
on-coplanar points on $P^1(C)$ (in terms of the extended hyperbolic 3-space)
determines a unique ideal (or totally asymptotic) tetrahedron of finite
invariant volume (by using the constant negative curvature of hyperbolic
3-space).  Since the orientation preserving isometry group is $PSL(2,C)$,
we can take 3 of the 4 vertices to be $\infty $, 0, 1, the 4-th point is
then defined to be the ``cross-ratio'' of the 4 distinct points (which may
determine a degenerate tetrahedron when they are coplanar).  (2.1b) is
the result of taking 5 distinct points: $\infty $, 0, 1, $x$ and $y$ as
pictured in (Fig. 1).  For a general division
ring $F$, $P_F$ merely formalizes the discussion.  The difference between
$P(F)$ and $P_F$ amounts to permitting some of the vertices to be duplicated.
(2.3) and (2.4) express the fact that oriented volume changes sign when the
exchange of two vertices reverses the orientation.  The equality
$P_C = P(C)$ simply means that the introduction of degenerate tetrahedra
with duplicated vertices does not make any difference (it does make a
difference in the case of $F = R)$.  With (2.3) in place, it is now evident
that (1.7) and (1.8) are directly related to (2.4) and (2.1b).  The problem
is that our explanation so far is based on $F = C$ while $L^{PS}$ dealt with
$F = R$.  This will be reviewed in the next section.  It should be noted
that the volume calculation makes perfectly good sense for tetrahedra with
vertices in the finite part of the hyperbolic 3-space.  It is known that
any such tetrahedron can be written in many different ways as a sum and
difference of ideal tetrahedra, see [DS1].  A general volume formula for a
tetrahedron is quite complicated.  However, the volume of an ideal
tetrahedron is quite simple.  It is given by the
{\it imaginary} part of the complexified Rogers' dilogarithm function (up to
normalization) evaluated at the cross-ratio.
\medskip
We end the present section by giving the structures and inter-relations of the
groups $P(F)$, $F = R$, $C$, $H$, with $R \subset  C \subset  H$.  The details
can be found in [DPS] and [Sa3].

$$
P(C) = P(C)^+ \oplus  P(C)^-. \leqno(2.6)
$$
\noindent
This is a $Q$-vector space direct sum in terms of its $\pm 1$ eigenspaces
under the action of complex conjugation.  Both summands have continuum
dimension.

$$
0 \rightarrow  Q/Z \rightarrow  P(R) \rightarrow  P(C)^+
\rightarrow  \Lambda ^2_Z(R/Z) \rightarrow  0 \hbox{ is exact. } \leqno(2.7)
$$
\noindent
$P(R)$ is the direct sum of $Q/Z$ and a $Q$-vector space of
continuum dimension.

$$
P(C)^+ \rightarrow  P(H) \rightarrow  0\hbox{ is exact and }\ P(H) \cong
\Lambda ^2_Z(R^+). \leqno(2.8)
$$
\noindent
The group $P(C)^-$ is the ``scissors congruence group'' in hyperbolic 3-space,
see [DS1].  The kernel of the homomorphism in (2.8) is related to the
``scissors congruence group modulo decomposables'' in spherical 3-space and is
conjecturally equal to it, see [DPS].  These results depend on algebraic
$K$-theory and use, in particular, a special case of Suslin's
celebrated solution of the conjecture of Lichtenbaum-Quillen, see [Su2].
\bigskip
{\bf {\S}3.  Rogers' Dilogarithm and Characteristic Classes.}
\medskip
As reviewed in preceding sections, there is a formal resemblance between the
Rogers' dilogarithm identities and volume calculation in hyperbolic 3-space.
In fact, the underlying space is quite different.  The explanations were
carried out in [D1].  For the convenience of the reader, we review the results.
The relevant characteristic class is that of Cheeger-Chern-Simons
characteristic class $\hat c_2$ which lies in the third cohomology of
$SL(2,C)$ viewed as a {\it discrete} group and where the coefficients
lie in $C/Z$.  In general, one has
$\hat c_n$ which lies in the (2n-1)-th cohomology of $GL(m,C)$, $m \geq  n$,
viewed as a discrete group, where the coefficients lie in $C/Z$.  The
standard mathematical notation for this cohomology group is
$H^{2n-1}(BGL(m,C)^\delta ,C/Z)$, this
is the group cohomology where $GL(m,C)$ is given the discrete topology (the
superscript $\delta $ emphasizes this fact).  $\hat c_1$ is nothing more than
the logarithm of the determinant map with kernel $SL(m,C)$.  With the
replacement of $GL$ by $SL$, $\hat c_1$ becomes 0.  The replacement of
$GL(m,C)$ by $GL(n,C)$ arises from homological stability theorems, see [Su1]
(a simplified version can be found in [Sa2]).  In general, $\hat c_n$
is conjectured to be connected to the $n$-polylogarithm, see
[D2 and D3].  Although we are only interested in $\hat c_2$, we will state the
results for general $n$.  The construction arises by starting with the Chern
form $c_n$ (a $2n$-form) which represents an integral cohomology class of the
classifying space $BGL(n,C)$ where $GL(n,C)$ is now given the usual topology.
Since we have replaced the usual topology by the discrete topology (this
amounts to ``zero curvature condition''), it follows from Chern-Weil theory
(where closed forms are viewed as complex cohomology classes) that $c_n$ can be
written as the differential of a (2n-1)-form, (for $n = 2$ this is the
Chern-Simons form that appears ubiquitously in physics).  When the coefficients
are taken in $C/Z$, this (2n-1)-form is closed and leads to the class
$\hat c_n$ in $H^{2n-1}(BGL(n,C)^\delta ,C/Z)$ through the exact sequence:
$$
0 \rightarrow  Z \rightarrow  C \rightarrow  C/Z \rightarrow  0. \leqno(3.1)
$$
We now concentrate on $n = 2$.  If we take the coefficients to be $C/Z$, then
the characteristic class $\hat c_2$ has a purely imaginary
part and a real part.  The purely imaginary part has values in $R$ and is
related to volume calculation in hyperbolic 3 space while the real part lies in
$R/Z$ and is related to volume calculations in spherical 3-space.  These volume
calculations are classically known to involve the dilogarithm function.  See
[C] for the details related to the work of Lobatchevskii and Schl\"afli
respectively.  The integer ambiguity in the spherical case arises from the fact
that a large tetrahedron can be viewed as a small tetrahedron on the ``back
side'' of the sphere with a reversed orientation.  Thus its volume is only
unique up to an integer multiple of the total volume of the spherical 3-space.
\medskip
For the Rogers' dilogarithm, the space is actually the group-space
$\widetilde{S}$ of the universal covering group $\widetilde{PSL(2,R)}$.  The
task of defining a tetrahedron and calculating its volume becomes more
delicate.  If we select a base point $p$ in $\widetilde{S}$, then any
point can be written as $g(p)$ for a uniquely determined group
element $g$ of $\widetilde{PSL(2,R)}$.  We first define a left invariant
``geodesic'' in the group that joins 1 to $g$ (this definition is
asymmetric).  This can be accomplished by exponentiating a Cartan
decomposition of the Lie algebra of $\widetilde{PSL(2,R)}$.  In essence,
we coordinatize $\widetilde{PSL(2,R)}$ by $R \times  H^2$ where $H^2$
denotes the hyperbolic plane.  Inductively, we can then define a ``geodesic
cone'' for any ordered set of $n+1$ points, $n \geq  0$, see (Fig. 2).
This is similar to [GM] where Rogers' dilogarithm appeared in terms of volumes
in Grassmann manifolds of 2-planes in $R^4$.  Our interpretation is dual to
[GM] since the transpose of a $4 \times  2$ matrix is a $2 \times  4$ matrix.
 Namely, for the ordered set $(p_0,\cdot \cdot \cdot ,p_n)$, the cone is the
collection of all points on the ``geodesics'' from $p_0$ to the ``geodesic
cone'' inductively defined for $(p_1,\cdot \cdot \cdot ,p_n)$.  For the
definition of volume $(n = 3)$, the next step is to show that it is enough to
consider the case where the
4 vertices are close to each other.  In fact, in terms of the Cartan
coordinates of the group elements, one may assume that the
$\theta $-coordinates are strictly positive and small (this involves changing
by a boundary which causes no problem because the volume is obtained by
evaluating a 3-cocycle on the chain, in essence we invoke Stoke's
Theorem).  We next form the boundary $R \times  \partial \ H^2$ where
$\partial \ H^2 = P^1(R)$ is the projective line over the real numbers (which
can be identified with $\{-\infty \} \cup  R$ by using the slopes in the right
half plane as in [PS]).  At this point, we begin to mimic the hyperbolic
3-space and move $p$ continuously towards $\{0\} \times  P^1(R)$ (this amounts
to right multiplication).  When $p$ lands on $\{0\} \times  P^1(R)$, so will
all four vertices so that we have the analog of an ideal hyperbolic
tetrahedron.  The volume (up to a normalizing factor) is just the value of the
Rogers' dilogarithm evaluated on the ``cross ratio'' of the ordered set of
vertices viewed as points of $P^1(R)$ (adjustments are needed for the
degenerate cases).  The situation now resembles the case of spherical 3-space.
Namely, the final volume will involve an integer (after normalization)
ambiguity which depends on the path of $p$.  We ignore the question of
representing the original tetrahedron as sums and differences of these ``ideal
tetrahedra'' since our concern is to interpret the value of the Rogers'
dilogarithm as a volume.

\bigskip
\centerline{(Fig. 2)}
\bigskip
We summarize this discussion in the form, cf. [D1, Th. 1.11]:
\medskip
{\bf Theorem 3.2.}  The restriction of the second Cheeger-Chern-Simons
characteristic class $\hat c_2$ to $PSL(2,R)$ can be lifted to a real
cohomology class on the universal covering group $\widetilde{PSL(2,R)}$ and
is then given by the Rogers' dilogarithm (more precisely, by $L^{PS}$
through $L$).
\medskip
\noindent
A more detailed discussion will be given in the following sections.
\bigskip
{\bf {\S}4.  Homology of Abstract Groups.}
\medskip
The basic reference is [Br].  Let $G$ be an abstract group.  We consider the
non-homogeneous formulation of the integral homology of $G$ with integer
coefficients $Z$.  The $j-th$ chain group $C_j(G)$ is the free abelian group
generated by all $j$-tuples $[g_1|\cdot \cdot \cdot |g_j]$ with $g_i$ ranging
over $G$, $j \geq  1$.  $C_0(G)$ is the infinite cyclic group generated by
$[\cdot ]$.  Such a $j$-cell should be identified with each of the formal
$j$-simplices
$(g_0,g_0g_1,g_0g_1g_2,\cdot \cdot \cdot ,g_0g_1\cdot \cdot \cdot g_j)$ as
$g_0$ ranges over $G$.  The boundary homomorphism: $\partial _j: C_j(G)
\rightarrow  C_{j-1}(G)$ is defined by translating the usual boundary of the
formal $j$-simplex.  For example, $\partial _3[g_1|g_2|g_3] = [g_2|g_3] -
[g_1g_2|g_3] + [g_1|g_2g_3] - [g_1|g_2]$.  The $j-th$ integral homology group
of $G$, $H_j(G,Z)$, or simply $H_j(G)$, is defined to be
$\ker \ \partial _j/{\rm im} \ \partial _{j+1}$.  $H_0(G)$ is just $Z$ while
$H_1(G)$
is canonically the commutator quotient group of $G$ with the class of $[g]$
mapped onto the coset of $g$ in the commutator quotient group.  We note that
homology groups can also be defined for any $G$-module $M$ (e.g. any vector
space on which $G$ acts by means of linear transformations).  This
generalization is often needed for computational purposes and requires more
care.
\medskip
In general, the procedure described in the preceding paragraph is not very
revealing.  Somewhat more revealing is to use the action of $G$ of a suitably
selected set $X$.  Typically, we end up describing the homology groups through
a spectral sequence that reveals a composition series.  If $X$ is the
underlying set of $G$ under the left multiplication action and the spectral
sequence ``degenerates''.  In the case of $\widetilde{PSL(2,R)}$, we can take
the space
$X$ to be that of $P^1(R) = \{0\} \times  P^1(R)$ which is viewed as part of
the boundary of the group space $\widetilde{S}$.  The spectral sequence is the
algebraic
procedure to keep track of the geometry.  If $p$ is a base point in the group
space $\widetilde{S}$, the 3-cell $[g_1|g_2|g_3]$ is an abstraction of the
``geodesic''
3-simplex $(p,g_1(p),g_1g_2(p),g_1g_2g_3(p))$ in the group space
$\widetilde{S}$.  If $p$
is moved to $\infty  = R(^1_0)$ in $P^1(R)$, then we have an ``ideal''
3-simplex.  Although the action of $\widetilde{PSL(2,R)}$ on $\widetilde{S}$ is
faithful, its
action on $P^1(R)$ is not.  In fact, it factors through $PSL(2,R)$ by way of
the following exact sequence:

$$
0 \rightarrow  Z\cdot c \rightarrow  \widetilde{PSL(2,R)} \rightarrow
PSL(2,R) \rightarrow  1. \leqno(4.1)
$$
\medskip
\noindent
The results in [PS] and [DPS] can be recast and summed up by the following
commutative diagram of maps where the rows and columns are exact:

$$

%
\matrix{%
&&&&& 0 && 0 && \cr
&&&&& \mapdown{} && \mapdown{} && \cr
&&&&& Z & \mapright{\cong} & Z \cdot c && \cr
&&&&& \mapdown{} && \mapdown{} && \cr
&&& 0 & \mapright{} & H_3(\tilde S, Z) & \mapright{} & PS(R) && \cr
&&&&& \mapdown{} && \mapdown{\eta} && \cr
& 0 & \mapright{} & Z_2 & \mapright{} & H_3(S,Z) & \mapright{\sigma} & P(R) &
\mapright{d^2} & \Lambda^2 (R^+) \cr
&&&&& \mapdown{} && \mapdown{} && \cr
&&&&& 0 && 0 $$ \cr
} \leqno(4.2)
$$
\medskip
\noindent
In (4.2), we abuse the notation and set $S = PSL(2,R)$.  $PS(R)$ is the abelian
group generated by all cross-ratio symbols
$\{r\} = (\infty ,0,1,r)$, $r \in  R^\times  \cup
\{\infty \}$, and subjected to the defining relations, cf. (1.5), (1.8):

$$
\{r_1\} - \{r_2\} + \{{r_2 \over r_1}\} - \{{r_2-1 \over r_1-1}\} +
\{{1-r^{-1}_2 \over 1-r^{-1}_1} \} = 0, 1 < r_1 < r_2, \leqno(4.3)
$$
$$
\{r\} + \{r^{-1}\} = 0\hbox{, } r > 1, \leqno(4.4)
$$
$$
\{\infty \} = 2\{2\} = -2\{1/2\}\hbox{ and } \{1\} = 0, \leqno(4.5)
$$
$$
\{-r\} = \{1+r^{-1}\} + \{\infty \}\hbox{, } r > 0.  \leqno(4.6)
$$

\noindent
These involve slight modifications of the results in [PS].  The group
$PS(R)$ is isomorphic to the group $H_3(W/S)$ of [PS] if we simply view
(4.4) through (4.6) as the definition of $\{s\}$ for $0 < s < 1$, $s = \infty $
or 1 and $s < 0$ respectively.  More precisely, we take as $j$-cells the
{\it ordered} $(j+1)$-tuples of elements of the universal covering group $R$
of $PSO(2,R)$ so that the convex closure of these points cover an interval of
length less than $\pi$ (length of $PSO(2,R)$).  Moreover, we also enlarge
the action to the "universal covering group" of $PGL(2,R)$.  We note that
in general, the universal covering group of a disconnected Lie group is not
well defined.  In the present case, it is well defined and happens to be
a semi-direct product of the universal covering group of $PSL(2,R)$ by an
element of order $2$ that inverts its infinite cyclic center.  The later
results in [DPS] and [Sa3] showed that $H_3(W/S)$ is a $Q$-vector space.
In [PS], it was shown that $H_3(W/S)/Z\cdot 48\{2\} \supset  H_3(SL(2,R),Z)$
and $H_3(W/S)/Z\cdot 12\{2\} \cong  P_R \supset  H_3(PSL(2,R),Z)$.
The first arose by showing that a certain element $c(-1,-1) = 8c$ is
mapped onto $\pm 48\{2\}$ (with a little care, the image is $-48\{2\})$.
The second involves a direct argument.  We note that $H_3(SL(2,R),Z)$ maps
surjectively to $H_3(PSL(2,R),Z)$ with kernel $Z_4$.  This accounts for
various $Z_2$'s. (4.2) now results from (2.5) with $c$ mapped by $\eta$
onto -6[[2]] in $P(R)$, namely, $P(R) \cong  H_3(W/S)/Z\cdot 6\{2\}$.
{}From section 1, we have a surjective homomorphism:

$$
L^{PS} : PS(R) \rightarrow  R, \hbox{ where } L^{PS}(\{s\}) = L(s)
- {\pi ^2 \over 6} = -L(1-s), 0 < s \leq  1. \leqno(4.7)
$$
\medskip
\noindent
In particular, $L^{PS}(\{1/2\}) = -\pi ^2/12$ and $L^{PS}(\{r\}) =
L(1-r^{-1})$, for $1 \leq  r \leq  \infty .$
\medskip
\noindent
This leads to surjective homomorphisms:

$$
\eqalign {L^{PS}_R &: P_R \rightarrow R \hbox{ mod } Z \cdot (\pi ^2)
\cr L^{PS}(R) &: P(R) \rightarrow  R \hbox{ mod } Z \cdot ({\pi ^2 \over 2}).
\cr}
\leqno(4.8)
$$
\medskip
\noindent
Using (2.5) and (2.7) we then have:

$$
\eqalign {L^{PS}_R &: H_3(PSL(2,R),Z) \rightarrow  R \hbox{ mod } Z \cdot (\pi
^2) \cr
L^{PS}(R) &: H_3(PSL(2,R),Z) \rightarrow  R \hbox{ mod } Z \cdot ({\pi ^2 \over
2}). \cr} \leqno(4.9)
$$
\medskip
\noindent
$L^{PS}_R$ is injective on torsion elements and $L^{PS}(R)$ maps an element of
order $m$ to one of order $m$ or $m/2$ according
to $m$ is odd or even.
\medskip

{\bf Remarks 4.10.}  (i) In using the extension to $PGL(2,R)$ and its
universal covering group, $[[r]]$ is the usual cross-ratio symbol associated
to $(\infty ,0,1,r)$ for $r$ in $R -\{0,1\}$, see [PS].  Thus, $\{r\}$ is
mapped to $[[r]]$.  (ii) $H_3(PSL(2,R),Z)$ is conjectured to be equal to
$H_3(PSL(2,R^{alg}),Z)$ where $R^{alg}$ denote the field of all real
algebraic numbers.  This follows from a similar conjecture for $C$ in place
of $R$.  Thus, the two maps in (4.9) are not expected to be surjective.  So
far, all the non-trivial elements in the image are obtained by using
algebraic numbers.  (iii) It is both convenient and essential to consider
the group $H_3(PSL(2,C),Z)$ or $H_3(SL(2,C),Z)$.  Namely, $C$ admits a huge
group of automorphisms while $R$ has only the trivial
automorphism.  While we do not know the injectivity of $\hat c_2 :
H_3(SL(2,C),Z) \rightarrow  C/Z $, we do know that a non-zero element
of $H_3(SL(2,C^{alg}),Z)$ can be detected by a composition
$\hat c_2 \circ  \tau $
for a suitable automorphism $\tau $ of $C$.  This is a theorem of Borel, see
[Bo].  Except when $\tau $ is the identity or the complex conjugation map, the
image $\tau (R)$ is everywhere dense in $C$.  It is the use of the hyperbolic
volume interpretation that ultimately leads to conclusion that $H_3(SL(2,C),Z)$
and $H_3(SL(2,R),Z)$ both contain a $Q$-vector subspace of infinite dimension.
\bigskip
{\bf {\S}5.  Connection with Richmond-Szekeres and Kirillov-Reshetikhin
Identities.}
\medskip
Granting the assertions in the preceding reviews, we can now describe the
relation of the above discussions with the Richmond-Szekeres identity (1.9) and
the extension by Kirillov-Reshetikhin (1.10).  As described in [PS], if $G$ is
a cyclic group of order $m$ with generator $g$, then the following chain is a
$(2j-1)$-cycle and its class generates $H_{2j-1}(G,Z) \cong  Z_m$, $j > 0:$

$$
c^{(j)}_m = \sum [g|x_1|g|\cdot \cdot \cdot |x_{j-1}|g]\hbox{, } x_i\hbox{
range over }\ G \hbox{ independently.} \leqno(5.1)
$$
\medskip
\noindent
More generally, $\sum [g^{i(1)}|x_1|\cdot \cdot \cdot |x_{j-1}|g^{i(j)}]$ is
homologous to $i(1)\cdot \cdot \cdot i(j)\cdot c^{(j)}_m$.  The superscript is
used to remind us that the class behaves as a $j-th$ power character on the
cyclic groups.  We now map $G$ into $S = PSL(2,R)$ by sending $g$ to the
following matrix:

$$
\left[  ^{\cos \ \theta   \ \ -\sin \ \theta }_{\sin \ \theta \ \ \ \cos \
\theta }
\right] \hbox{, }  \theta  = \pi /m \hbox{.}
$$
\medskip
\noindent
The map $\sigma $ in (4.2) sending $H_3(S,Z)$ into $P(R)$ is obtained by
sending the 3-cell $[g_1|g_2|g_3]$ to the cross-ratio symbol of
$(\infty ,g_1(\infty ),g_1g_2(\infty ),g_1g_2g_3(\infty ))$.  Here $\infty  =
R(^0_1)$, $r = R(^1_r)$, more generally, $y/x = R(^x_y)$, $x \geq  0$ and
$PGL(2,R)$ acts on these lines through matrix multiplication.  However, as
discussed in section 3, in the evaluation of volume, chains may be modified
by boundaries.  For the special form of
the 3-cells that appears in $c^{(2)}_m$, this is not a serious problem.
In any event, we have a canonical identification of the torsion subgroup:

$$
tor(H_3(PSL(2,R),Z)) \cong  Q\pi /Z\pi \hbox{, the rational rotations
in }\ PSO(2,R). \leqno(5.2)
$$
\medskip
We now consider $c_m = c^{(2)}_m$ and note that $\sigma (c_m)$ is of order $m$
or $m/2$ in $P(R)$ according to $m$ is odd or even.  Thus, we will restrict
ourselves to $m > 2$.  $[g|g^j|g]$ corresponds to
$(\infty ,g(\infty ),g^{j+1}(\infty ),g^{j+2}(\infty ))$.  Except when $j = 0$,
$m-2$, $m-1$, this is just $[[Q^2_j/Q_{j-1}Q_{j+1}]]$, where $Q_j =
Q_j(\theta ) = \sin (j+1)\theta / \sin \theta $, $\theta  = \pi /m.$
\medskip
When $j = 0$.  $[g|1|g]$ is 0 under the usual normalization.  The corresponding
formal 3-cell has two identical adjacent vertices and represents 0.
\medskip
When $j = m-2 > 0$.  We have the formal 3-cell $(\infty ,-1,1,\infty )$
independent of $m$.  It is the same as $(\infty ,0,1,\infty )$ and is assigned
the cross ratio symbol $\{\infty \}$.  By taking the boundary of
$(\infty ,0,1,2,\infty )$, $\{\infty \}$ is homologous to $2\{2\} =
-2\{1/2\}$ as in (4.5).
\medskip
When $j = m-1 \geq  2$.  We have the formal 3-cell $(\infty ,0,\infty ,0)$
independent of $m$.  It is the boundary of $(\infty ,0,\infty ,0,1)$.  Thus,
we set it to 0.
\medskip
To see how the preceding assignments work, we consider the cases: $m = 3$ and
4.
\medskip
When $m = 3$,  $\sigma (c_3) = [[\infty ]]$ and $L^{PS}(\{\infty \}) =
\pi ^2/6$.  This represents an element of order 3 in $R$ mod
$Z\cdot (\pi ^2/2)$.
\medskip
When $m = 4$, $\sigma (c_4) = [[\infty ]] + [[2]]$ and $L^{PS}(\{\infty \}) +
L^{PS}(\{2\}) = \pi ^2/6 + \pi ^2/12 = \pi ^2/4$ .  This represents an element
of order 2 in $R$ mod $Z(\pi ^2/2).$
\medskip
We now go to the general case.  For $m > 2$, we have:

$$
\eqalign {\sigma (c_m) &= [[\infty ]] + \sum _{1 \leq  j \leq m-3}[[{Q^2_j
\over Q_{j-1}Q_{j+1} }]]\hbox{, } \cr
Q_j = Q_j(\theta ) &= {\sin (j+1)\theta \over \sin\theta} \hbox{, } 1 \leq  j
\leq  m-3\hbox{, } \theta  = \pi /m. \cr} \leqno(5.3)
$$
\medskip
\noindent
The above calculation is purely formal and the only reason that $\theta $ is
chosen to be $\pi /m$ arises from the fact that the expression in (5.1)
represents the image of an element of order $m$ or $m/2$ in $H_3(S,Z)$.  The
expression for $Q_j$ is well known in terms of representation theory.  Namely,
consider the irreducible representations of $SL(2,C)$ of finite dimension.  It
is well known that there is exactly one in each dimension $n+1 \geq  1$.  It is
realized in the $n-th$ symmetric powers of the fundamental representation of
$SL(2,C)$ on $C^2$.  This is the spin $n/2$ representation in physics.
Evidently, the matrix diag$(z,z^{-1})$ is represented by
diag$(z^n,z^{n-2},\cdot \cdot \cdot ,z^{-n})$.  $Q_j(\theta )$ is just the
trace of diag$(z,z^{-1})$ in the spin $j/2$ representation where $z =
\exp (\iota \theta )$.  The following lemma results from looking at the
character of the representation theory of $SL(2,C):$
\medskip
{\bf Lemma 5.4.}  Let $S(i)$ denote the $i-th$ symmetric tensor representation
of $SL(2,C)$, $i \geq 0$.  Let $j, p, q > 0$.  Then $S(p+j-1) \otimes  S(q+j-1)
\cong  S(p-1) \otimes  S(q-1) \oplus  S(p+q+j-1) \otimes S(j-1)$ holds.
(Note: the representation $S(i)$ has degree $i+1$.)
\medskip
For the proof, it is enough to looke at the trace of the matrix
$diag(z,z^{-1})$.  If we consider the special case of $z = \exp (\iota \theta
), p = q = 1$,
we get $Q^2_i = Q_{i-1}Q_{i+1} + 1$.  Since $Q^2_j = 1/d_j$ by definition, we
have:
$$
\sigma (c_m) = [[\infty ]] + \sum _{1 \leq j \leq m-3}[[(1-d_j)^{-1}]].
\leqno(5.5)
$$
\medskip
\noindent
The right hand side of (5.5) is $[[\infty ]] +
2\ \sum _{1 \leq j \leq k-1}[[(1-d_j)^{-1}]]$ for $m = 2k+1$ and is
$[[\infty ]] + [[(1-d_k)^{-1}]] + 2\ \sum _{1 \leq j \leq k-1}[[1-d_j)^{-1}]]$
for $m = 2k+2$.
\medskip
We next have the following elementary result:
\medskip
{\bf Lemma 5.6.}  Let $F: Q \rightarrow  Q$ be an additive homomorphism so
that $F(Z) \subset  Z$ and so that $F: Q/Z \cong  Q/Z$.  Then $F = \pm Id$.  If
$F(1/3) \equiv  -1/3$ mod $Z$, then $F = -Id.$
\medskip
Proof.  Recall that $F$ is just multiplication by a rational number because
division by integers is unique.  The two restrictions on $F$ force $F$ to be
multiplication $by \pm 1$.  The final restriction forces $F$ to be minus
identity.
\medskip
We can now apply Lemma 5.6 to obtain the following:
\medskip
{\bf Theorem 5.7.}  For $m \geq  3 \hbox{, } L^{PS}(R) (\sigma (c_m)) \equiv
-\pi ^2/m \hbox{ mod } Z \cdot (\pi ^2/2)$.  In general, we have the congruence
Kirillov-Reshetikhin identity:

$$
\sum _{1 \leq j \leq m-2}L({{\rm sin^2} {\pi \over m} \over {{\rm sin^2}
{(j+1)\pi \over m}}}) = {\pi ^2 \over 6} \cdot {3(m-2) \over m} \equiv
{-\pi ^2 \over m} \hbox{ mod } Z \cdot ({\pi ^2 \over 2}).
$$
In particular, we have the congruence Richmond-Szekeres identity for m = 2k+1:

$$
\sum _{1 \leq j \leq k-1}L({{\rm sin^2} {\pi \over 2k+1} \over {\rm sin^2}
{(j+1)\pi \over 2k+1}}) \equiv {\pi^2 (2k-2) \over 6(2k+1)} \hbox{ mod } Z
\cdot
({\pi^2 \over 4}).
$$

Proof.  We already know that $Q\{2\}$ is the inverse image of the torsion
subgroup of $P(R)$ in $PS(R)$.  Moreover, $L^{PS} : Q\{2\} \rightarrow
Q\pi ^2$ is an isomorphism that carries $6\{2\}$ onto $\pi ^2/2$.  The torsion
subgroup of $H_3(PSL(2,R),Z)$ is identified with $Q\pi /Z\pi $ where the
elements $c_m$ arising from rotation by $\pi /m$ in $PSO(2,R)$ and
$\sigma (c_m)$ has order $m$ or $m/2$ in $P(R)$ according to $m$ is odd or
even.  Since $c_m$ corresponds to $\pi /m$ in $Q\pi /Z\pi $, Lemma 5.6 shows
that $L^{PS}(R)(\sigma (c_m))$ must $be \pm \pi ^2/m$ in $Q\pi ^2$ mod
$Z\cdot (\pi ^2/2)$.  When $m = 3$, we saw that the image is $\pi ^2/6 =
\pi ^2/2 - \pi ^2/3$.  It follows that $L^{PS}(R)(\sigma (c_m)) = -\pi ^2/m$
mod $Z\cdot (\pi ^2/2)$.  This is just the general congruence identity.  The
more precise equality was proved in [KR-II, (2.33) and Appendix 2.] by an
analytic argument.
\medskip
Let $m = 2k+1$.  By (4.4), (4.7), and $\sin  (\pi  - \phi ) = \sin  \phi $,
$L^{PS}(R)(\sigma (c_m)) = \pi ^2/6 + 2\sum _jL(d_j)$, $1 \leq  j \leq  k-1$.
Next $\pi ^2/2 - \pi ^2/(2k+1) = (2k-1)\pi ^2/2(2k+1) = \pi ^2/6 +
(4k-4)\pi ^2/6(2k+1)$. The congruence immediately follows.
\medskip
If we use the fact $L^{PS}$ is injective on $Q\{2\}$, we have the corollary:
\medskip
{\bf Corollary 5.8.}  In $PS(R), 4(m-3)\{2\} = m \cdot \sum _{1 \leq j \leq
m-3} \{(1-d_j)^{-1}\}, m > 2.$
Equivalently, $6(m-2)\{2\} = m \cdot \sum _{1 \leq j \leq m-2 }
\{(1-d_m)^{-1}\}, m > 2.$
\medskip
We may obtain more congruence identities by computing the image in $PS(R)$ of
a representative for the class $p \cdot q \cdot c_m^{(2)}$, $0 < p \hbox{, } q
< m$.  Namely, we take $i(1) = p$ and $i(2) = q$ in the extension of (5.1).
There are at most 4 exceptional symbols to consider according to $j$ mod $m$.
When $j = 0$, we always have 0.  We therefore assume $0 < j < m$.  If
$j = -p$ or $-q$, depending on $p = q$ or $p \neq q$, we end up with either
0 or $- \{\infty\}$.  Finally, if $j \equiv -p-q$ mod $m$ (this forces
$p + q \neq m$), then the symbol is $\{\infty\}$ as before.  The general
congreunce identity then takes on the folloing form:

\medskip
{\bf Theorem 5.9.}  Let $L^{PS}$ denote the shifted Rogers' dilogarithm as in
(4.7).  Let $m > 2, 0 < p, q < m$.  Let

$$
\delta _j (p,q;m) = {{{\rm sin} (p+j)\theta \cdot {\rm sin} (q+j)\theta} \over
{{\rm sin} j\theta \cdot {\rm sin} (p+q+j)\theta}}, 0 < j < m, \theta = {\pi
\over m}.
$$

We then have the following congruence with the understanding that: the index
$j$ is to skip over the cases, $-p, -q, -p-q$ mod $m$; and $\delta _{a,b}$ is
the Kronecker delta mod $m$:

$$
\sum _{1 \leq j \leq m-1}L^{PS}(\{\delta_j (p,q;m)\}) \equiv -{pq\pi ^2 \over
m} + (\delta _{p,-q} - \delta _{p,q}) \cdot {\pi ^2 \over 6} \hbox{ mod } Z
\cdot ({\pi ^2 \over 2}).
$$
\medskip
\noindent
We note that the number $\delta _j(p,q;m)$ lies in $R - {0,1}$ after we exclude
the exceptional cases.  It is easy to see that ${{\rm sin} (x+p)\theta \over
{\rm sin} x}$ is strictly decreasing in $x$.  Thus, $\delta _j(p,q;m)$ can be
negative.  In general, it is necessary to use the defining properties
(4.3)-(4.7) of $L^{PS}$ in order to express the congruence in terms of $L$.
If we use Lemma 5.4, it is easy to see that:
$$
\delta_j(p,q;m) ^{-1} = 1 - {{\rm sin} p\theta \cdot {\rm sin} q\theta \over
{\rm sin} (p+j)\theta \cdot {\rm sin} (q+j)\theta}.
$$
In the case of $p = q = 1$, the right hand side is strictly between 0 and 1
so that (4.4) and (4.7) recover the congruence in Theorem 5.7.  However, for
general $p, q$, we do not have a good way to determine the "integral
ambiguity" implicit in lifting the congruence to an identity.  This resembles
the classical treatment of Gauss' quadratic reciprocity theorem in number
theory via the use of Gauss' sums.

\medskip

{\bf Remark 5.10.}  In Theorem 5.7, the rational numbers:  $(2k-2)/(2k+1)$ are
the ``so-called'' effective central charge of the $(2,2k+1)$ non-unitary
Virasoro minimal model.  Similarly, the rational number $3 \ell /(\ell + 2)$
is the central charge of the level $\ell \ A_1 ^{(1)} \ WZW$ model.  Both are
models in conformal field theory.  In our present setting, they are
identified as specific values of the evaluation of the Cheeger-Chern-Simons
characteristic class on the third integral homology of the universal covering
group $\widetilde{PSL(2,R)}$ of $PSL(2,R)$ (viewed as a discrete group).  These
homology classes are the lifts of the torsion classes for $PSL(2,R)$.
\medskip
In the recent work of Kirillov [K] concerning a conjecture of Nahm on the
spectrum of rational conformal field theory [NRT], the following abelian
subgroup $W$ of $Q$ was considered:

$$
W = \{\sum _i n_iL(a_i)/L(1)|\ n_i \in  Z\hbox{, } a_i \in  R^{alg}\} \cap  Q.
$$
\medskip
\noindent
{}From our discussion, it is clear that $W$ contains both 1 as well as $-1/m$
mod
$Z$ for every positive integer $m$.  Thus, $W$ is simply $Q$.  In the
conjecture of Nahm, one is more concerned with the set of effective central
charges and $n_i$ is assumed to be non-negative.  This is closed under addition
because one can form tensor product of models.  Our discussion only pins down
the fractional part of such central charges while the integral parts apparently
spread the central charges out in a way that resembled the volume distribution
of hyperbolic 3-manifolds.  In the present approach, these effective central
charges are volumes of certain 3-cycles in a totally different space--the
compactification of the universal covering group of $PSL(2,R)$.  These 3-cycles
can be viewed as ``orbifolds'' since they arise from the finite cyclic
subgroups of $SL(2,R)$.  It should also be noted that the central charge of
the Virasoro algebra is the value of a degree two cohomology class while
our description is on the level of degree three group cohomology, but for the
Lie groups viewed as discrete groups.  The precise relation between these
cohomologies is not too well understood.  On the level of classifying spaces of
topological groups, there is the wellknown conjecture, see [M] and [FM]:
\medskip
{\bf Conjecture of Friedlander-Milnor.}  Let $G$ be any Lie group and let $p$
be a prime.  Then $H_i(BG^\delta ,Z_p) \rightarrow  H_i(BG,Z_p)$ is an
isomorphism (it is known to be surjective).
\bigskip
{\bf {\S}6.  The ``beta map'' and various conjectures.}
\medskip
In the work of Nahm-Recknagel-Terhoeven, [NRT], speculations were made about
the relevance of algebraic $K$-theory, Bloch groups [Bl], geometry of
hyperbolic 3-manifolds [Th1] as well as the ``physical meaning'' of a
``beta map''.  To some extent, we have clarified the first three of these.
Namely, a connection between the effective central charge in rational
conformal field theory with algebraic $K$-theory and Bloch groups [Bl] can be
made by way of the second characteristic class of Cheeger-Chern-Simons and
its interpretation via volume calculation in the universal covering group
space of $PSL(2,R)$.  Specifically, it is {\it not} connected with the volume
calculation in hyperbolic 3-space.  (Note:  According to Thurston's work,
[Th], volume of hyperbolic 3-manifolds is a topological invariant.)  Roughly
speaking, the difference rests with a missing factor of $(-1)^{1/2}$.  We
next clarify the origin of the ``beta map''.  In terms of diagram (4.2),
the ``beta map'' is denoted by:

$$
d^2: P(R) \rightarrow  \Lambda ^2_Z(R^\times )\hbox{, } d^2([[r]]) = r \wedge
(r-1)\hbox{, } r > 1. \leqno(6.1)
$$
\medskip
\noindent
$d^2$ arises as the second differential in a spectral sequence.  It is defined
by solving a ``descent equation''.  This is typical of the higher differential
maps in a spectral sequence.  The exactness of the rows in (4.2) showed that
$\ker \ d^2 = {\rm im} \sigma $.  If we move up to the level of $PS(R)$, it
is then clear that the vanishing of the $d^2$-invariant characterizes the
elements of $H_3(\widetilde{PSL(2,R)},Z)$.  The origin of $d^2$ comes from the
Dehn invariant in Euclidean 3-space.  In 1900, Dehn used it to solve Hilbert's
Third Problem and extended it to hyperbolic and spherical 3-space, see [DS2].
By working with $P(C)$, see [DS1] and [DPS], $d^2$ then incorporates both
versions of the Dehn invariants.  In the present case, we would interpret
$d^2$ in terms of ``ideal polyhedra'' in $\widetilde{S}$.  As pointed out
in [PS], the following conjecture is still open:
\medskip
{\bf Conjecture 6.2.}  $L^{PS}: H_3(\widetilde{PSL(2,R)},Z) \rightarrow  R$ is
injective.
\medskip
\noindent
We already mentioned the following conjecture along this line:
\medskip
{\bf Conjecture 6.3.}  $H_3(\widetilde{PSL(2,R^{alg})},Z) \rightarrow
H_3(\widetilde{PSL(2,R)},Z)$ is
bijective.
\medskip
\noindent
The preceding conjecture is a special case of the more general ``folklore''
conjecture:
\medskip
{\bf Conjecture 6.4.}  $H_3(SL(2,C^{alg}),Z) \rightarrow  H_3(SL(2,C),Z)$ is
bijective.
\medskip
\noindent
More precisely, Conjecture 6.3 is equivalent to any of the corresponding
conjecture for a nontrivial quotient group of $\widetilde{PSL(2,R)}$, for
example
$PSL(2,R)$.  $H_3(SL(2,R),Z)$ is known to be isomorphic to the fixed point set
of $H_3(SL(2,C),Z)$, see [Sa3].  The map in Conjecture 6.4 is known to be
injective, see [Su2].  Thus Conjectures 6.3 and 6.4 would follow from:
\medskip
{\bf Conjecture 6.5.}  $H_3(SL(2,C^{alg}),Z) \rightarrow  H_3(SL(2,C),Z)$ is
surjective.
\medskip
\noindent
It should be mentioned that the map $H_3(SU(2),Z) \rightarrow  H_3(SL(2,C),Z)$
has image equal to the image of $H_3(SL(2,R),Z)$.  In this connection, we have:
\medskip
{\bf Conjecture 6.6.}  $H_3(SU(2),Z) \rightarrow  H_3(SL(2,C),Z)$ is injective.
\medskip
{\bf Conjecture 6.7.}  $\hat c_2 : H_3(SL(2,C),Z) \rightarrow  C/Z. $ is
injective.
\medskip
\noindent
Conjecture 6.7 is equivalent to the conjunction of conjecture 6.6 and the
converse of the Hilbert's Third Problem for hyperbolic as well as
spherical polytopes in dimension 3.  Namely, the Dehn invariant together with
volume detect the scissors congruence classes of such polytopes.  The Euclidean
case was solved by Dehn-Sydler, see [DS2] for discussions.  The best result
in this direction is the theorem of Borel, [Bo]:
\medskip
{\bf Borel's Theorem.}  Suppose $c$ is non-zero in $H_3(SL(2,C^{alg}),Z)$,
then $\hat c_2(\tau (c))$ is non-zero for a suitable automorphism $\tau $ of
$C.$
\medskip
We note that an illustration of the idea behind Borel's Theorem was the
proof given in [PS] that $H_3(SL(2,R^{alg}),Z)$ contains a rational vector
space of infinite dimension.  Recall, we consider a real algebra number $r_p$
satisfying the equation $X^p - X + 1 = 0$, $p$ an odd prime.  $d^2(\{r_p\})$ is
therefore 0 and $[[r_p]]$ then defines an element of $H_3(SL(2,R^{alg}),Z)$.
Since $L^{PS}$ is strictly monotone, there is no problem showing that we have
distinct elements.  However, it is not obvious that these elements are
$Q$-linearly independent.  This stronger statement was a combination of Galois
theory together with the use of the hyperbolic volume.

\bigskip
{\bf {\S}7.  Concluding Remarks.}
\medskip
In the present work, we showed that the effective central charges
for certain models in conformal field theory can be connected to the
evaluation of a real valued cohomology class on a suitable degree 3
homology class for the integral group homology of the universal
covering group $\widetilde {PSL(2,R)}$ of $PSL(2,R)$.  The important
point is that we have replaced the usual topology by the discrete
topology.  In addition, instead of the hyperbolic 3-space, we use
the group space of this universal covering group.  The particular
homology class is a suitable lift of a
homology class of finite order that generates the third integral
homology of a finite cyclic subgroup of $PSL(2,R)$.  The lift is
connected with the Rogers' dilogarithm identities due to Richmond-Szekeres
[RS] and Kirillov-Reshetikhin [KR].  All these identities are shown
to originate from the basic identities found by Rogers [R].  Our route
ends in the central charge identification but there are no firm
connections between any of
the intermediate steps followed by us with the intermediate steps
used in solvable models in conformal field theory.  A casual
reading of [BPZ] and [Z2] does show the many appearances of cross-ratios.
However, instead of the complex numbers or the real numbers, we see
meromorphic functions.  This is also the basic theme in the work of
Bloch [Bl].  On the mathematical side, there are efforts to build up
enormous structures to explain the steps used in the physics side.
Our present effort does not do this.

	Another of the principal points in the present work is the fact
that Rogers' dilogarithm has long been known to be connected with the
second Cheeger-Chern-Simons characteristic class which is represented
by the Chern-Simons form that appears in many current theoretical physics
investigations.  This connection is related to the interplay between
the "continuous" picture and the "discrete" picture.  On the mathematics
side, we have a direct map on the level of classifying spaces for groups
equipped with two topologies: one discrete, the other continuous.  The
map is the one that goes from the discrete to the continuous.  On the
physics side, the passage from the discrete to the continuous is a
subject of debate since there does not appear to be a specific
map (in the mathematical sense).  However, there are still a large
number of unresolved issues on the mathematical side.  For example, the
Virasoro algebra is typically viewed as the algebraic substitute for the
diffeomorphism group of the circle.  (More precisely, it may be viewed
as the ``pseudo-group'' of holomorphic maps on the sphere with two
punctures).  This contains $PSL(2,R)$ which acts as a group of
diffeomorphisms on the circle through the identifiction of the circle
with $P^1 (R)$.)  Our procedure replaces these infinite dimensional
(pseudo-) groups by the finite dimensional subgroups.
However, it is also accompanied by the use of the discrete topology.
Although the process of playing off one topology against another is
familiar in foliation theory, it is not explored in the present work.

	In passing, we would like to indicate that Rogers' dilogarithm
has appeared in various related works on the physics side.  Aside from
the work [BR] that led Bazhanov to ask one of us (CHS) about the
connection between [BR] and [PS] in the summer of 1986-7, there are the
earlier works of Zamolodchikov [Z2] and Baxter [B].  Specifically, in
the appendix of [B], Rogers dilogarithm appeared.  This has been
extended recently in [BB] where they have shown that the 3-d models
of Zamolodchikov can be related to the earlier 2-d chiral Potts models
considered in [AMPTY], [MPST], and [BPA] after suitable generalizations.
On the mathematics side, Atiyah and Murray [A] have identified the
algebraic curves in [BPA] and [MPST] as the spectral curves of N
magnetic monopoles arranged cyclically around an axis in hyperbolic
3-space.  In view of the fact that our present work indicates that
the group manifold $\widetilde {PSL(2,R)}$ is more appropriate than
the hyperbolic 3-space, one can not help but ask if there might be an
interesting mathematical theory of magnetic monopoles in
$\widetilde {PSL(2,R)}$.  Evidently, the present work raises many
more questions than it answers.

\bigskip
{\bf Acknowledgements:}  It is evident that we have benefited from many, many
colleagues.  Since it is impractical to list them all, we will, at
the risk of insulting many, limit ourselves to Profs. B. M. McCoy, C. N. Yang,
I. M. Gelfand and L. Takhtajan for inspiring tutorial discussions.  Special
thanks are due to J. D. Stasheff and V. V. Bazhanov for raising the crucial
questions at the right time, these questions led to the present work.

\vfill\eject
\bigskip
\centerline{References}
\bigskip
\item{[ABF]} G. E. Andrews, R. J. Baxter, and P. J. Forrester, Eight vertex
SOS model and generalized Rogers-Ramanujan identities, J. Stat. Phys. 35
(1984) 193-266.

\item{[A]}  M. F. Atiyah, Magnetic monopoles and the Yang-Baxter equations,
Int. J. Mod.   Phys. A 6 (1991) 2761-2774.

\item{[AMPTY]} H. Au-Yang, B. M. McCoy, J. H. H. Perk, S. Tang, and M.-L. Yan,
Commuting tranfer matrices in the chiral Potts models:  Solutions of
star-triangle equations with genus $> 1$, Phys. Lett. A, 123 (1987) 219-223.

\item{[B]}  R. J. Baxter, The Yang-Baxter equations and the Zamolodchikov
model,  Physica 18D (1986) 321-347.

\item{[BPA]} R. J. Baxter, J. H. H. Perk, and H. Au-Yang, New solutions of the
star triangle relations for the chiral Potts model, Phys. Lett. A 128 (1988),
138-142.

\item{[BB]}  V. V. Bazhanov and R. J. Baxter:

\itemitem{1.} New solvable lattice models in three dimensions, J. Stat. Phys.
69 (1992) 453-485.

\itemitem{2.} Star-Triangle relation for a three dimensional model,
(preprint, hep-th/9212050).

\item{[BR]}  V. V. Bazhanov and N. Yu Reshetikhin:
\itemitem{1.} Critical RSOS models and conformal
field theory, Int. J. Mod. Phys. A (1989) 115-142.
\itemitem{2.} RSOS models connected with simply laced algebras and
conformal field theory, J. Phys. A: Math. Gen. 23 (1990) 1477-1492.

\item{[BPZ]} A. A. Belavin, A. M. Polyakov, A. B. Zamolodchikov, Infinite
conformal symmetry in two-dimensional quantum field theory, Nucl. Phys. $B\ 241
(1984)$ 3-38.

\item{[Bl]}  S. Bloch, Higher regulators, algebraic $K$-theory, and zeta
functions of elliptic   curves, Irvine Lecture Notes, 1978.

\item{[Bo]}  A. Borel, Cohomologie de $SL_n$ et valeurs de fonctions zeta aux
points entiers,   Ann. Scien. Norm. Super. Pisa, Cl. Sci., 4 (1977) 613-636,
Correction, ibid. 7   (1980) 373.

\item{[Br]}  K. S. Brown, Cohomology of Groups, Springer Verlag, New York,
1982.

\item{[C]}  H. S. M. Coxeter, The functions of Schl\"afli and Lobatschefsky,
Quar. J. Math. 6 (1935) 13-29.

\item{[DKKMM]} S. Dasmahapatra, R. Kedem, T. R. Klassen, B. M. McCoy, and
E. Melzer, Quasi-particles, conformal field theory, and $q$-series, (preprint,
hep-th/9303013).

\item{[D]}  J. L. Dupont:
\itemitem{1.} The dilogarithm as a characteristic class for flat
bundles, J. Pure   App. Algebra, 44 (1987) 137-164.
\itemitem{2.} On polylogarithms, Nagoya Math. J. (1989) 1-20.
\itemitem{3.}  Characteristic classes for flat bundles and their
formulas, (preprint, Aarhus Univ. no. 23, 1992).

\item{[DPS]} J. L. Dupont, W. Parry, and C. H. Sah, Homology of classical Lie
groups made   discrete.  II.  $H_2$, $H_3$, and relations with scissors
congruences, J. Algebra, (1988)   215-260.

\item{[DS]}  J. L. Dupont and C. H. Sah, Scissors congruences, II:
\itemitem{1.} J. Pure App. Algebra 25   (1982) 159-195.
\itemitem{2.} Homology of Euclidean groups of motions made discrete, Acta
Math. 164 (1990) 1-27.

\item{[FS]}  E. Frenkel and A. Szenes, Dilogarithm identities, $q$-difference
equations and the   Virasoro algebra, (preprint, hep-th/9212094).

\item{[FM]}  E. M. Friedlander and G. Mislin, Cohomology of classifying spaces
of complex Lie groups and related discrete groups, Comm. Math. Helv. 59 (1984)
7-36.

\item{[GM]}  I. M. Gelfand and R. D. MacPherson, Geometry in Grassmannians
and a generalization of the dilogarithm, Adv. Math. 44 (1982) 279-312.

\item{[KKMM]} R. Kedem, T. R. Klassen, B. M. McCoy, and E. Melzer:
\itemitem{1.} Fermionic sum representations for conformal field theory
characters, (preprint, hep-th/9211102).
\itemitem{2.} Fermionic quasi-particle representations for characters of
$(G^{(1)})_1 \times  (G^{(1)})_1/(G^{(1)})_2$, (preprint, hep-th/9301046).

\item{[K]}  A. N. Kirillov:
\itemitem{1.} Identities for the Rogers dilogarithm function connected
with   simple Lie algebras, J. Sov. Math. 47 (1989) 2450-2459.
\itemitem{2.} Dilogarithm identities and spectra in conformal field
theory,   (preprint, \break\hfill
hep-th/9211137, also preprint, hep-th/9212150).
\item{[KR]}  A. N. Kirillov and Yu. N. Reshetikhin, Exact solutions of the
integrable $XXZ$ Heisenberg model with arbitrary spin: I, II, J. Phys. A: Math.
Gen. 20 (1987)   1565-1585, 1587-1597.

\item{[KP]}  A. Kl\"umper and P. A. Pearce, Conformal weights of RSOS lattice
models and   their fusion hierarchies, Physica A 183 (1992) 304-350.

\item{[KN]}  A. Kuniba and T. Nakanishi:
\itemitem{1.} Spectra in conformal field theories from the Rogers dilogarithm,
(preprint, hep-th/9206034).
\itemitem{2.} Rogers dilogarithm in integrable systems, (preprint,
hep-th/9210025).

\item{[KNS]} A. Kuniba, T. Nakanishi and J. Suzuki, Characters in conformal
field theories   from Thermodynamic Bethe Ansatz, (preprint, hep-th/9301018).

\item{[MPST]} B. M. McCoy, J. H. H. Perk, S. Tang and C. H. Sah, Commuting
transfer matrices for the 4-state self-dual Potts model with a genus 3
uniformizing  Fermat curve, Phys. Lett. A 125 (1987) 9-14.

\item{[M]}  J. W. Milnor, On the homology of Lie groups made discrete, Comm.
Math. Helv. 58 (1983) 72-85.

\item{[NRT]} W. Nahm, A. Recknagel and M. Terhoeven, Dilogarithm identities in
conformal   field theory, (preprint, hep-th/9211034).

\item{[PS]}  W. Parry and C. H. Sah, Third homology of $SL(2,R)$ made discrete,
J. Pure   App. Algebra 30 (1983) 181-209.

\item{[RS]}  B. Richmond and G. Szekeres, Some formulas related to
dilogarithms, the zeta   function and the Andrews-Gordon identities,
J. Austral. Math. Soc. 31 (1981)   362-373.

\item{[R]}  L. J. Rogers, On function sum theorem connected with the series
$\sum ^\infty _1 x^n/n^2$,   Proc. London Math. Soc. 4 (1907) 169-189.

\item{[Sa]}  C. H. Sah:
\itemitem{1.} Scissors congruences, I.  The Gauss-Bonnet map, Math. Scand.
49   (1981) 181-210.
\itemitem{2.} Homology of Lie groups made discrete, I.  Stability theorems
and   Schur multipliers, Comm. Math. Helv. 61 (1986) 308-347.
\itemitem{3.}  Homology of Lie groups made discrete, III, J. Pure and App.
Algebra   56 (1989) 269-312.

\item{[Su]}  A. A. Suslin:
\itemitem{1.} Homology of $GL_n$, characteristic classes and Milnor
$K$-theory,   Springer Lec. Notes in Math. 1046 (1984) 357-375.
\itemitem{2.} Algebraic $K$-theory of fields, Proc. Int. Cong. Math.,
Berkeley, I   (1986) 222-224.

\item{[Te]}  M. Terhoeven, Lift of dilogarithm to partition identities,
(preprint, hep-th/9211120).

\item{[Th]}  W. P. Thurston:
\itemitem{1.} The geometry and topology of three-manifolds.  (Lecture
Notes, Princeton, 1978-9).
\itemitem{2.} Three-Dimensional Geometry and Topology.  (Preprint, Notes,
Minnesota, 1990-1).

\item{[V]}  A. Varchenko, Multidimensional hypergeometric functions in
conformal field   theory, algebraic $K$-theory, algebraic geometry, Proc. Int.
Cong. Math., Kyoto, I   (1990) 281-300.

\item{[Z]}  A. B. Zamolodchikov:
\itemitem{1.} Tetrahedron equations and the relativistic $S$-Matrix of
straight-strings in $2+1$ dimensions, Comm. Math. Phys. 79 (1981) 489-505.
\itemitem{2.} Infinite additional symmetries in two-dimensional
conformal   quantum field theory, Theor. Math. Phys. 63 (1985) 1205-1213.
\vfill\eject\end